\documentstyle[12pt,german]{article}
\begin{document}
\begin{flushright}
CERN-TH.7236/94
\end{flushright}
\vspace*{5mm}
\begin{center}
{\bf The Breaking of Subnuclear Democracy\\
\bigskip
as the\\
\bigskip
Origin of Flavour Mixing}\\
\vspace*{6mm}
HARALD FRITZSCH$^{*+}$\\
\vspace*{3mm}
\it {Theory Division, CERN, CH-1211 Geneva 23, Switzerland}\\
\bigskip
\it and\\
\bigskip
\it {Max--Planck--Institut f\"ur Physik und Astrophysik}\\
\it {Werner--Heisenberg--Institut f\"ur Physik, Munich}\\
\vspace*{3mm}
{\rm and}\\
\vspace*{3mm}
{\rm DIRK HOLTMANNSP\"OTTER}\\
\vspace*{3mm}
\it {Sektion Physik, Universit\"at M\"unchen}\\
\end{center}
\vspace*{1cm}
\begin{center}
{\bf Abstract}\\
\end{center}
It is shown that the simplest breaking of the subnuclear democracy leads
to
a successful description of the mixing between the second and third
family.
In the lepton channel the $\nu _{\mu } - \nu_{\tau }$ oscillations are
expected to be described by a mixing angle of $2.65^ {\circ }$ which
might
be observed soon in neutrino experiments.\\

\bigskip
\rule{30mm}{0.2mm}\\
$^*$ {\footnotesize On leave from Sektion Physik, Universit\"at
M\"unchen}\\
$^+$ {\footnotesize Supported in part by DFG-contract 412/22--1}\\
\vspace*{1.5cm}
\begin{flushleft}
CERN-TH.7236/94 \\
April 1994
\end{flushleft}
\newpage
\setcounter{page}{1}
\pagestyle{plain}
\noindent
In the standard electroweak model both the masses of the quarks as well
as
the weak mixing angles enter as free parameters. Any further insight
into
the yet unknown dynamics of mass generation would imply a step beyond
the
 physics
of the electroweak standard model. At present it seems far too early to
attempt an actual solution of the dynamics of mass generation, and one
is
invited to follow a strategy similar to the one which led eventually to
the
solution of the strong interaction dynamics by QCD, by looking for
specific patterns and symmetries as well as specific symmetry
violations.\\
\\
The mass spectra of the quarks are dominated strongly by the masses of
the
members of the third family, i.\ e.\ by $t$ and $b$. Thus a clear
hierarchical
pattern exists. Furthermore the masses of the first family are small
compared
to those of the second one. Moreover, the CKM--mixing matrix exhibits a
hierarchical pattern -- the transitions between the second and third
family
as well as between the first and the third family are small compared to
those between the first and the second family.\\
\\
About 15 years ago, it was emphasized$^{1)}$ that the observed
hierarchies
 signify
that nature seems to be close to the so--called ``rank--one'' limit, in
which
all mixing angles vanish and both the u-- and d--type mass matrices are
proportional to the rank-one matrix\\
\bigskip
\begin{equation}
M_0 = {\rm const.} \cdot \left(\begin{array}{ccc}
0 & 0 & 0\\ 0 & 0 & 0\\ 0 & 0 & 1 \end{array} \right) \, .
\end{equation}
\\
\\
\noindent
Whether the dynamics of the mass generation allows that this limit can
be
achieved in a consistent way remains an unsolved issue. Encouraged by
the
observed hierarchical pattern of the masses and the mixing parameters,
we
shall assume that this is the case. In itself it is a non-trivial
constraint
and can be derived from imposing a chiral symmetry, as emphasized in
ref. (2).
This symmetry ensures that an electroweak doublet which is massless
remains
unmixed and is coupled to the $W$--boson with full strength. As soon as
mass is introduced, at least for one member of the doublet, the symmetry
is
violated and mixing phenomena are expected to show up. That way a chiral
evolution of the CKM matrix can be considered.$^{2)}$ At the first stage
only the $t$ and $b$ quark masses are introduced, due to their
non-vanishing
coupling to the scalar ``Higgs'' field. The CKM--matrix is unity in this
limit. At the next stage the second generation acquires a mass also.
Since
the $(u, d)$--doublet is still massless, only the second and the third
 generations
mix, and the CKM--matrix is given by a real $2 \times 2$ rotation matrix
in the
$(c, \, s) - (t, \, b)$ subsystem, describing e.\ g.\ the mixing between
$s$
and $b$. Only at the next step, at which the $u$ and $d$ masses are
introduced, does the full CKM--matrix appear, described in general by
three angles
 and one
phase.\\
\newpage
\noindent
It has been emphasized some time ago$^{3)}$ that the rank-one mass
matrix (see
 eq.
(1)) can be expressed in terms of a ``democratic mass matrix'':\\
\bigskip
\begin{equation}
M_0 = c \left( \begin{array}{ccc}
1 & 1 & 1\\ 1 & 1 & 1\\ 1 & 1 & 1 \end{array} \right) \, ,
\end{equation}
\\
\\
which exhibits an $S(3)_L \, \, \times \, \, S(3)_R$ symmetry. Writing
down
the mass eigenstates in terms of the eigenstates of the
``democratic'' symmetry, one finds e.g. for the lepton channel:\\
\begin{eqnarray}
e^0 & = & \frac{1}{\sqrt{2}} (l_1 - l_2) \nonumber\\
\mu^0 & = & \frac{1}{\sqrt{6}} (l_1 + l_2 - 2l_3)\\
\tau^0 & = & \frac{1}{\sqrt{3}} (l_1 + l_2 + l_3) \nonumber
\end{eqnarray}
\\
($l_i$: symmetry eigenstates). Note that $e^0$ and $\mu^0$ are massless
in
the limit considered here, and any linear combination of the first two
state vectors given in eq. (3) would fulfil the same purpose, i.\ e.\
the
decomposition is not unique, only the wave function of the coherent
state
$\tau^0$ is uniquely defined. This ambiguity will disappear as soon as
the symmetry is violated.\\
\\
The wave functions given in eq. (3) are reminiscent of the wave
functions
of the neutral pseudoscalar mesons in QCD in the $SU(3)_L \, \, \times
\, \,
SU(3)_R$ limit:\\
\begin{eqnarray}
\pi^0_0 & = & \frac{1}{\sqrt{2}}(\bar uu - \bar dd)\\
\eta_0 & = & \frac{1}{\sqrt{6}}(\bar uu + \bar dd - 2\bar ss)
\nonumber\\
\eta '_0 & = & \frac{1}{\sqrt{3}}(\bar uu + \bar dd + \bar ss) .
\nonumber
\end{eqnarray}
\\
(Here the lower index denotes that we are considering the chiral limit).
Also the mass spectrum of these mesons is identical to the mass spectrum
of
the leptons and quarks in the ``democratic'' limit: two mesons $(\pi
^0_0 \, ,
\eta _0)$ are massless and act as Nambu--Goldstone bosons, while the
third
coherent state $\eta '_0$ is \underline{not} massless due to the QCD
anomaly.\\
\newpage
\noindent
In the chiral limit the (mass)$^2$--matrix of the neutral pseudoscalar
mesons
is also a ``democratic'' mass matrix when written in terms of the $(\bar
qq)$--
eigenstates $(\bar uu), \, (\bar dd)$ and $(\bar ss)^{4)}$:\\
\bigskip
\begin{equation}
M^2(ps) = \lambda \left( \begin{array}{ccc}
1 & 1 & 1\\ 1 & 1 & 1\\ 1 & 1 & 1 \end{array} \right),
\end{equation}
\\
\\
where the strength parameter $\lambda $ is given by
$\lambda = M^{2}(\eta '_{0}) \, / \, 3$.
The mass matrix (5) describes the result of the QCD--anomaly which
causes strong
transitions between the quark eigenstates (due to gluonic annihilation
effects
enhanced by topological effects). Likewise one may argue that analogous
transitions are the reason for
the lepton--quark mass hierarchy. Here we shall not speculate about a
detailed mechanism of this type, but merely study the effect of symmetry
breaking.\\
\\
In the case of the pseudoscalar mesons the breaking of the symmetry down
to
$SU(2)_L \, \, \times \, \, SU(2)_R$ is provided by a direct mass term
$m_s \bar
 ss$
for the s--quark. This implies a modifica\-tion of the (3,3) matrix
element in
eq. (5), where $\lambda $ is replaced by $\lambda + M^2(\bar ss)$ where
 $M^2(\bar ss)$ is
given by $2M^2_k$, which is proportional to $< \bar ss >_0$, the
expectation
value of $\bar ss$ in the QCD vacuum. This direct mass term causes the
violation of the symmetry and generates at the same time a mixing
between
$\eta _0$ and $\eta '_{0}$, a mass for the $\eta _{0}$, and a mass shift
for
the $\eta '_{0}$.\\
\\
It would be interesting to see whether an analogue of the simplest
violation of the ``democratic'' symmetry which describes successfully
the
mass and mixing pattern of the $\eta - \eta '$--system is also able to
describe the observed mixing and mass pattern of the second and third
family of leptons and quarks. Let us replace the (3,3) matrix element in
eq. (2) by $1 + \varepsilon _i$; (i = l (lepton), u (u--quark), d
(d--quark)
respectively. The small real parameter $\varepsilon $ describes the
departure
from democratic symmetry and leads\\
\\
a) \hspace*{0.2cm} \parbox[t]{15cm}
{to a generation of mass for the second family and}\\
\\
b) \hspace*{0.2cm} \parbox[t]{15cm}
{to flavour mixing between the third and the second
family. Since $\varepsilon $ is directly related (see below) to a
fermion mass
and the latter is \underline{not} restricted to be positive,
$\varepsilon $
can be positive or negative. (Note that a negative Fermi--Dirac mass can
always be turned into a positive one by a suitable $\gamma
_5$--transformation
of the spin $\frac{1}{2}$ field). Since the original mass term is
represented by a symmetric matrix, we take $\varepsilon $ to be real.}\\
\newpage
\noindent
First we study the mass and mixing pattern of the charged leptons. The
mass
operator (trace $\Theta^{\mu} _{\mu}$ of the energy--momentum tensor
$\Theta_{\mu \nu})$ can be written as\\
\bigskip
\begin{equation}
\Theta ^{\mu }_{\mu } = \Theta ^{0 \mu }_{\mu } + c_l \varepsilon _l
\bar l_3 l_3
\end{equation}
\\
\\
where $\Theta ^{0 \mu}_\mu $ describes the mass term in the symmetry
limit.
The modification of the spectrum and the induced mixing can be obtained
by
considering the matrix elements:\\
\begin{eqnarray}
< \mu ^0 | c_l \varepsilon _l \bar l_3 l_3 | \mu ^0 > & = & +
\frac{2}{3} c_l
\varepsilon _l \nonumber\\
< \tau ^0 | c_l \varepsilon _l \bar l_3 l_3 | \tau ^0 > & = & +
\frac{1}{3}
c_l \varepsilon _l\\
< \mu ^0 | c_l \varepsilon _l \bar l_3 l_3 | \tau ^0 > & = & -
\frac{\sqrt{2}}
{3} c_l \varepsilon _l \nonumber \, \, .
\end{eqnarray}
\\
\\
One observes that\\
\\
a) \hspace*{0.2cm} \parbox[t]{15cm}{the muon acquires a mass given by
$c_l
\cdot \varepsilon _l$ i.\ e.\ $m(\mu ) / m(\tau ) \cong \frac{2}{9}
\varepsilon _l$;}\\
\\
b) \hspace*{0.2cm} \parbox[t]{15cm}{the $\tau $--lepton mass is changed
slightly
($m(\tau ) / m(\tau ^0) \cong 1 + \frac{1}{9} \varepsilon _l)$;}\\
\\
c) \hspace*{0.2cm} \parbox[t]{15cm}{the flavour mixing is induced by the
fact
that the perturbation proportional to $\bar l _3 l_3$ leads to a
non--vanishing
transition matrix element between $\mu ^0$ und $\tau ^0$.}\\
\\
\\
This phenomenon is
analogous to the chiral symmetry violation of QCD, where the s--quark
mass
term $m_s \bar ss$ leads to a mass for the $\eta $--meson, a mass shift
for
the $\eta '$--meson and a mixing between $\eta $ and $\eta '$.\\
\\
It is instructive to rewrite the mass matrix in the hierarchical basis,
where
one obtains, using the relations (7):\\
\bigskip
\begin{equation}
M = c_l \left( \begin{array}{ccc}
0 & 0 & 0\\ 0 & + \frac{2}{3}{\varepsilon _l} & - \frac{\sqrt{2}}{3}
\varepsilon
 _l\\
0 & - \frac{\sqrt{2}}{3} \varepsilon _l & 3 + \frac{1}{3}\varepsilon _l
 \end{array}
\right) \, .
\end{equation}
\newpage
\noindent
In lowest order of $\varepsilon $ one finds the mass eigenvalues
$m_\mu = \frac{2}{9} \varepsilon _l \cdot m_\tau \, , m_\tau = m_{\tau
^0} \, ,
\Theta _{\mu \tau} = | \sqrt{2} \cdot \varepsilon _l / 9|$.\\
\\
The exact mass eigenvalues and the mixing angle are given by:\\

\begin{eqnarray}
m_1 / c_l & = & \frac{3 + \varepsilon _l}{2} - \frac{3}{2} \sqrt{1 -
\frac{2}{9} \varepsilon _l + \frac{1}{9} \varepsilon _l ^2} \nonumber\\
m_2 / c_l & = & \frac{3 + \varepsilon _l}{2} + \frac{3}{2} \sqrt{1 -
\frac{2}{9}
\varepsilon _l + \frac{1}{9} \varepsilon _l^2}\\
\sin \Theta _l & = & \frac{1}{\sqrt{2}}\left( 1 - \frac{1 -
 \frac{1}{9}\varepsilon _l}
{(1 - \frac{2}{9} \varepsilon _l + \frac{1}{9} \varepsilon
 ^2_l)^{1/2}}\right)^{1/2} \nonumber
\end{eqnarray}
\\
\\
The ratio $m_{\mu } / m_{\tau }$, observed to be $0.0595$, gives
$\varepsilon _l = 0.286$ and a $\mu - \tau $ mixing angle of
$2.65^{\circ}$.
Whether this mixing angle is directly relevant for neutrino oscillations
or
not depends on the neutrino sector. For massless neutrinos the mixing
angle
does not have a direct physical meaning, i.\ e.\ it can be rotated away.
If
neutrinos have a mass, the neutrino mass matrix will in general induce
further
mixing angles. A general discussion would be beyond the scope of this
paper.\\
\\
However, we should like to consider an interesting scenario which is
being
discussed in connection with cosmological aspects. Let us suppose that
the $\tau $--neutrino mass is of the order of 10 eV in order to be
relevant for
the ``missing matter problem'' in cosmology, the muon neutrino is in the
milli--eV range, i.\ e.\ $m(\nu _{\mu }) < 10^{-2} eV$, and the electron
neutrino mass is neglected. The mass generation for the $\nu _{\mu
}$--mass
proceeds in an analogous way as discussed above for the muon mass.
However,
the $\varepsilon $--parameter for the neutrino sector is tiny $(< 5
\cdot
 10^{-2})$,
and the mixing angle induced via the $\nu _{\mu }$--mass generation can
safely be neglected. Thus the angle relevant for the
$\nu _{\mu } - \nu _{\tau }$ oscillations remains
$2.65^{\circ }$, i.\ e.\ $\sin^{2} 2\Theta = 0.0085$.
This value is essentially the lowest limit given by the Charm II
 experiment$^{5)}$,
i.\ e.\ is not ruled out for any value of
$\bigtriangleup m^2 = m(\nu _{\tau })^2 - m(\nu _{\mu })^2$.
However, the E531 experiment$^{6)}$ gives a limit of about $16 eV^2$ for
$\bigtriangleup m^2$, i.\ e.\ $m(\nu _{\tau }) < 4 eV$. This limit seems
to
rule out a cosmological role with respect to the ``missing matter'' for
the
$\tau $--neutrino. However, one might caution this conclusion since our
mixing angle of $2.65^{\circ }$ is not far from the limit of
$(\sin^2 2 \Theta = 0.004)$, at which, according to the E531 experiment,
all
values of $m(\nu _{\tau })$ are allowed. New experiments, e.\ g.\ the
CHORUS and NOMAD experiments now or soon under way at CERN, will clarify
this
 issue.
If the mixing angle is $2.65^{\circ }$ as argued above and the $\nu
_{\tau
 }$--mass
above 10 eV, one should observe the $\nu _{\mu } - \nu _{\tau }$--
oscillations within one year$^{7)}$.\\
\newpage
\noindent
Replacing $\varepsilon _l$ by $\varepsilon _u$, $\varepsilon _d$
respectively, we can determine the symmetry breaking parameters for the
quark sector. The ratio $m_s / m_b$ is allowed to vary in the range
$0.022 \ldots 0.044$ (see ref. (8)). According to eq. (9) one finds
$\varepsilon _u$ to vary from $\varepsilon _d = 0.11$ to $0.21$.
The associated $s - b$
mixing angle varies from $\Theta (s, b) = 1.0^{\circ }$ \hspace{0.3cm}
$(\sin \Theta = 0.018)$ and $\Theta (s, b) = 1.95^{\circ }$
\hspace{0.3cm}
$(\sin \Theta = 0.034)$. As an illustrative example we use the values
$m_b(1GeV)
 = 5200 MeV$, \hspace{0.3cm}
$m_s(1GeV) = 220 MeV$. One obtains $\varepsilon _d = 0.20$ and
$\sin \Theta(s, b) = 0.032$.\\
\\
To determine the amount of mixing in the $(c, t)$--channel, a knowledge
of
the ratio $m_c / m_t$ is required. As an illustrative example we take
$m_c(1GeV) = 1.35 GeV$, $m_t(1GeV) = 260 GeV$ (i.\ e.\ $m_t(m_t) =
160GeV)$,
which gives $m_c / m_t \cong 0.005$. In this case one finds $\varepsilon
_u =
 0.023$ and $\Theta(c,t) = 0.21^{\circ }$
\hspace{0.3cm} $(\sin \Theta (c, t) = 0.004$) .\\
\\
The actual weak mixing between the third and the second quark family is
combined effect of the two family mixings described above. The symmetry
breaking given by the $\varepsilon $--parameter can be interpreted, as
done
in eq. (7), as a direct mass term for the $l_3(u_3, d_3)$ fermion
system.
However, a direct fermion mass term need not be positive, since its sign
can always be changed by a suitable $\gamma _5$--transformation. What
counts
for our ana\-lysis is the relative sign of the $m_s$--mass term in
comparison
to the $m_c$--term, discussed previously. Thus two possibilities must be
considered:\\
\\
a) \hspace*{0.2cm} \parbox[t]{15cm}{Both the $m_s$-- and the $m_c$--term
have the same relative sign with respect to each other, i.\ e.\ both
 $\varepsilon _d$
and $\varepsilon _u$ are positive, and the mixing angle between the
second
and third family is given by the difference $\Theta (sb) - \Theta (ct)$.
This
possibility seems to be ruled out by experiment, since it would lead to
$V_{cb} < 0.03$.}\\
\\
\\
b) \hspace*{0.2cm} \parbox[t]{15cm}{The relative signs of the breaking
terms
$\varepsilon _d$ and $\varepsilon _u$ are different, and the mixing
angle
between the $(s,b)$ and $(c,t)$ systems is given by the sum
$\Theta(sb) + \Theta(ct)$. Thus we obtain $V_{cb} \cong \sin (\Theta(sb)
+ \Theta(ct))$.}\\
\\
\\
According to the range of values for $m_s$ discussed above, one finds
$V_{cb} \cong 0.022 ... 0.038$.
For example, for $m_s(1GeV) = 220MeV$, \, $m_c (1GeV) = 1.35 GeV$,
$m_t(1GeV)
= 260GeV$ one finds $V_{cb} \cong 0.036$.\\
\\
Before discussing the experimental situation, we add a comment about the
mass
ge\-neration for the first family, which at the same time will also
generate
the other mixing elements, e.g. $V_{us}$ and $V_{ub}$, of the CKM
matrix. These
masses can be generated by a further breakdown of the symmetry, e.\ g.\
in the
matrix of eq. (5) by a small departure of a second diagonal matrix
element
from unity. (This would correspond to a direct mass term for that
state.) Due to the small values of the masses of the first family in
comparison to the $\lambda $--scale, given by the mass of the third
generation fermion (e.g. $m_e / \lambda = 0.0009)$, the strength of this
symmetry breaking is much smaller than the primary symmetry breaking,
which
leads to the masses for the second family. (The situation is analogous
to the
one in hadronic physics, where the breaking of the chiral symmetry is
primarily
given by the mass of the s--quark, and the $m_u / m_d$ mass terms can
be neglected to a good approximation). In general it is expected, both
from the
 arguments considered here and more
generally from the analysis on chiral symmetry given in ref. (2), that
the
matrix elements $V_{cb}$ and $V_{ts}$ will be affected only by small
corrections
of order $10^{-3}$ or less in absolute magnitude (of order
$\frac{m_d}{m_b}$,
$\frac{m_u}{m_t}$ respectively). Thus the primary breaking of the
democratic symmetry leads solely to a mixing between the second and the
third
family, and the secondary breaking, responsible for the Cabibbo angle
etc.,
will not affect the $2 \times 2$ submatrix of the CKM--matrix describing
the $s
 - b$
mixing in a significant way.\\
\\
The experiments give $V_{cb} = 0.032 \dots 0.054^{9)}$. We conclude from
the analysis
 given above
that our ansatz for the symmetry breaking reproduces the lower part of
the
experimental range. According to a recent analysis the experimental data
are reproduced best for $V_{cb} = 0.038 \pm 0.003^{10)}$, i.\ e.\ it
seems
that $V_{cb}$ is lower than previously thought, consistent with our
expectation. Nevertheless we obtain consistency with experiment only if
the ration $m_s / m_b$ is relatively large implying $m_s(1GeV) \ge
180MeV$.\\
\\
It is remarkable that the simplest ansatz for the breaking of the
``democratic
symmetry'', one which nature follows in the case of the pseudoscalar
mesons, is able to reproduce the experimental data on the mixing between
the second and third family. We interpret this as a hint that the
eigenstates
of the symmetry $l_{i}, q_i$ respectively, and not the mass eigenstates,
play
a special r\^{o}le in the physics of flavour, a r\^{o}le
which needs to be investigated further.\\
\\
\underline{Acknowledgement}:\\
\\
One of us (H.F.) is indebted to Prof. K. Winter
for useful discussions on neutrino oscillations.\\

\newpage
\noindent
{\bf References}
\bigskip
\begin{enumerate}
\renewcommand{\baselinestretch}{1}
\item H. Fritzsch, Nucl. Phys. \underline{B155} (1979), 189;\\
See also: H. Fritzsch, in: Proc. Europhysics Conf. on Flavor Mixing,\\
Erice, Italy (1984).\\

\item H. Fritzsch, Phys. Lett. \underline{B184} (1987) 391.\\

\item H. Harari, H. Haut and J. Weyers, Phys. Lett. \underline{78B} (1978)
459;\\
Y. Chikashige, G. Gelmini, R.P. Peccei and M. Roncadelli,\\
Phys. Lett.\ \underline{94B} (1980) 499;\\
C. Jarlskog, in: Proc. of the Int. Symp. on Production and Decay of\\
Heavy Flavors, Heidelberg, Germany, 1986;\\
P. Kaus and S. Meshkov, Mod. Phys. Lett. \underline{A3} (1988) 1251;
\underline{A4} (1989) 603;\\
H. Fritzsch and J. Plankl, Phys. Lett. \underline{B237} (1990) 451.\\

\item H. Fritzsch and P. Minkowski, Nuovo Cimento \underline{30A} (1975)
393;\\
H. Fritzsch and D. Jackson, Phys. Lett. \underline{66B} (1977) 365.\\

\item M. Gruw\a'e et al., Phys.\ Lett.\ \underline{B309} (1993) 463.\\

\item N. Ushida et al., Phys. Rev. Lett.\ \underline{57} (1986) 2897.\\

\item K. Winter, private communication.\\

\item J. Gasser and H. Leutwyler, Physics Reports \underline{87} (1982)
77.\\

\item Particle Data Group, Review of Particle Properties,\\
Phys. Rev. \underline{D45} (1992) 1.\\

\item S. Stone, Syracuse preprint HEPSY 93-11 (1993).\\
\end{enumerate}
\end{document}